\documentclass[prl,twocolumn,showpacs]{revtex4}
\usepackage{graphicx}
\usepackage{amsmath}
\usepackage{amssymb}
\usepackage{epsf,latexsym}
\usepackage{times}
\newcommand{\ket}[1]{\left| #1 \right\rangle}

\newcommand{\be}{\begin{equation}}
\newcommand{\ee}{\end{equation}}
\newcommand{\ba}{\begin{eqnarray}}
\newcommand{\ea}{\end{eqnarray}}
\newcommand{\bc}{\begin{center}}
\newcommand{\ec}{\end{center}}

\begin{document}

\title{Creating and preserving multi-partite entanglement with spin chains}

\author{Irene D'Amico$^{1}$}
\email{ida500@york.ac.uk}
\author{Brendon W. Lovett$^{2}$}
\email{brendon.lovett@materials.oxford.ac.uk}
\author{Timothy P. Spiller$^{3}$}
\email{tim.spiller@hp.com}

\affiliation{
$^1$ Department of Physics, University of York, York YO10 5DD, United Kingdom\\
$^2$ Department of Materials, University of Oxford, OX1 3PH, United Kingdom\\
$^3$ Hewlett-Packard Laboratories, Filton Road,
Stoke Gifford, Bristol BS34 8QZ, United Kingdom}

\date{\today }

\begin{abstract}

We show how multi-partite entanglement, such as of W-state form,
can be created in branched spin chain systems. We also discuss
the preservation of such entanglement, once created.
The technique
could be applied to actual spin chain systems, or to other
physical systems such as strings of coupled quantum dots,
molecules or atoms.
\end{abstract}


%03.67.Lx Quantum computation
%78.67.Hc Quantum dots;78.67.-n Optical properties of low-dimensional, mesoscopic,
%and nanoscale materials and structures
%03.67.-a Quantum information
%73.21.La Quantum dots;73.21.-b Electron states and collective excitations in multilayers,
%quantum wells, mesoscopic, and nanoscale systems
%85.35.Be Quantum well devices (quantum dots, quantum wires, etc.);85.35.-p Nanoelectronic
%devices

\maketitle

% \Dsection{First section}

%%% 1 %%%

\section{Introduction}
The past few years has seen much interest in the
propagation of quantum information through spin chains. From a
practical quantum technology perspective, such controlled propagation
of quantum information has potential use. For macroscopic distances,
quantum states of
light form the best vehicle for quantum communication---these can
propagate through optical fibres or free space with high fidelity
\cite{qkd}. However, for much shorter distances it is
quite possible that other media, such as spin chains, could make a very
useful contribution. Future solid state quantum
information devices could require microscopic\\ quantum
communication links, between separate
quantum processors or registers, or between processors and memory,
analogous to the communication links that exist within the
computer chips of today.

We use the terminology ``spin chain'' for any
physical system that can be modelled using the Hamiltonian
of a coupled spin chain, introduced in the next section.
In reality, this system could be a chain of actual spins
(produced chemically, or fabricated) connected through
nearest-neigh-bour interactions,
or it could be a string of quantum dots or molecules (like
fullerenes), containing exciton or spin qubits. A string of trapped
atoms is another possibility.

To quantify how well quantum processors or registers might
communicate, there has been emphasis on the quality of quantum state
propagation through a spin chain or network. It has been
shown that a single spin qubit can transfer with decent fidelity
along a constant nearest-neigh-bour exchange-coupled chain
\cite{bose03}. The fidelity can approach unity for a qubit
encoded into a packet of spins \cite{osb04}. More complicated systems,
with different
geometry or chosen unequal couplings
\cite{chr04,chr05,dechi05,yun05,kar05} can also achieve perfect
state transfer, as can parallel spins chains \cite{bur05a,bur05b},
or a chain used as a
wire, with controlled couplings at the ends \cite{woj05}. Related
studies have been made on chains of coupled quantum dots
\cite{damico05}, or oscillators \cite{ple05,har05}
and spin chains connected through longer range magnetic dipole
interactions \cite{avel06}. State transfer and operations via
spin chains using adiabatic dark passage have also been considered
\cite{gre05,ohs07}.

In this work, rather than investigating state transfer, we consider
instead the approach of creating an entangled resource, using the
natural dynamics of a branched spin chain system and a
very simple initial state.
The consequences of branching were first studied in the context of
divided {\it bosonic} chains, composed of coupled harmonic
oscillators~\cite{plenio04,perales05}---and have
since been studied in systems of propagating electrons~\cite{yang}.
In both of
these works, the dynamics of Gaussian wave packet type excitations
were investigated. In contrast, our work considers the propagation of
a single spin-down excitation localized on a single site, and how this
moves through a network of spin-up states, generating entanglement.
Clearly such entanglement is a useful and flexible resource
\cite{gre05,plenio04,perales05,yang,dev05,pat05,spi06} for
quantum information. For example, it could be used for
quantum teleportation \cite{tele93}, which enables the transfer of a
quantum state, or to connect separated quantum registers or
processors.  One advantage of this approach is that decoherence or
imperfections, which can lead to imperfect entanglement production,
could be countered by purification techniques \cite{pur96} prior to
use of the entangled resource. This clearly requires some sort of
preservation of the entanglement, once created. We briefly discuss
this issue.

%%% 2 %%%

\section{Spin chain formalism}
Consider a one-dimensional
chain of $N$ spins, each coupled to their nearest neighbours. The
Hamiltonian for the system is
\begin{eqnarray}
%H &=& -\sum_{i=1}^N \frac{E_i}{2} \sigma_{z}^{i} \nonumber \\
%&+&
% \sum_{i=1}^{N-1} \frac{J_{i,i+1}}{2}\left(\sigma_{+}^{i}\sigma_{-}^{i+1} +
%\sigma_{-}^{i}\sigma_{+}^{i+1}\right)
% \;,
H &=& -\sum_{i=1}^N \frac{E_i}{2} \sigma_{z}^{i}
+\sum_{i=1}^{N-1} \frac{J_{i,i+1}}{2}\left(\sigma_{+}^{i}\sigma_{-}^{i+1} +
\sigma_{-}^{i}\sigma_{+}^{i+1}\right),
\label{Hspin}
\end{eqnarray}
where $\sigma_{z}^{i}$ is the $z$ Pauli spin matrix for the spin
at site $i$ and similarly $\sigma_{\pm}^i=\sigma_{x}^{i} \pm i
\sigma_{y}^{i}$. For actual spins, $E_{i}/2$ is the local magnetic
field (in the $z$-direction) at site $i$ and $J_{i,i+1}/2$ is the
local $XY$ coupling strength between neighbouring sites $i$ and
$i+1$. Alternatively, for a chain of quantum dots where each qubit is
represented by the presence or absence of a ground state exciton,
$E_i$ is the exciton energy and $J_{i,i+1}$ is the F\"{o}rster
coupling between dots $i$ and $i+1$ \cite{damico05}. The spin chain
formalism applies to both these physical systems, and to any others
that can be modelled by the Hamiltonian (\ref{Hspin}).
The computational basis notation for the spin
states at each site is $|0\rangle_i \equiv |\uparrow_z\rangle_i$ and
$|1\rangle_i \equiv |\downarrow_z\rangle_i$.

The total $z$-component of spin (magnetization), or total exciton
number, is a constant of motion as it commutes with $H$. It is
therefore useful to use a state of the system which is the ground
state, except for a single flipped spin. This state is
straightforward to prepare, assuming local control somewhere
(such as at the end of a chain) over a spin.
The flipped spin can be regarded as a
(conserved) travelling qubit as it moves around under the action of the chain
dynamics \cite{bose03}. An efficient way of representing
such states in an $N$ spin network is the site basis defined as
$|{\mathsf  k}\rangle = |{0_1, 0_2, ..., 0_{k-1}, 1_k, 0_{k+1}, ..., 0_N}\rangle$.
A system prepared in this subspace remains in it, with the detailed dynamics
depending on the local magnetic fields or exciton energies. However, if these
are independent of location $i$ then the dynamics favour quantum state
transfer processes, as already mentioned. We adopt this limit
in this work.

%%% 3 %%%

\section{Creation of multi-partite entanglement} In other works
\cite{ida07} we have considered the distribution and freezing
of bipartite entanglement, using a simple one-to-two branched spin
chain system and chosen couplings along the chain. A simple spin
flip at the end of the input branch can evolve into a superposition
of being at one output or the other, thus generating maximal
bipartite entanglement at the outputs. Furthermore, this
entanglement can be frozen against further dynamical evolution
if the final link of each output branch is an additional
one-to-two branch, and rapid phase flips are applied when the state
reaches the output \cite{ida07}. Such branched
systems can be analysed simply and elegantly using the results of
Christandl {\it et al.} in \cite{chr05}. They show that perfect transfer
along a spin chain is possible, using $N-1$ {\it unequal}
couplings for an $N$ site chain---between site $i$ and site $i+1$ given by
\begin{equation}
J_{i,i+1} = \alpha \sqrt{i(N-i)}\;.
\label{coupling}
\end{equation}
The  size of the interaction is set by the constant $\alpha$.
The authors of \cite{chr05} also demonstrate the projection of an array of spins
with  equal couplings $j$ onto a 1D chain of unequal couplings. For an array
arranged as a series of columns, with each node of column $i$ connected to
$n$ different nodes of adjacent column $i+1$, and each node of column $i+1$
connected to $m$ different nodes of column $i$, each column can be projected
onto a single node of a one-dimensional spin chain, with coupling $j\sqrt{mn}$
between nodes $i$ and $i+1$ \cite{chr05}. The consequence of this is that
at a one-to-two branching node, the two output couplings pick up
a factor of $1/\sqrt{2}$ compared to the value in the equivalent 1D chain
(which should satisfy (\ref{coupling}) for perfect transmission).
This same $1/\sqrt{2}$ branching factor also prevents hub reflection
for propagating wave packets in divided chains of
oscillators~\cite{perales05}, as well as Gaussian wave packets of
propagating electrons or magnons~\cite{yang}.

\begin{figure}[b]\bc%f5
\centering
%\vspace{-12.5cm}
\vspace{-11.5cm}
\includegraphics[width=1.5\columnwidth]{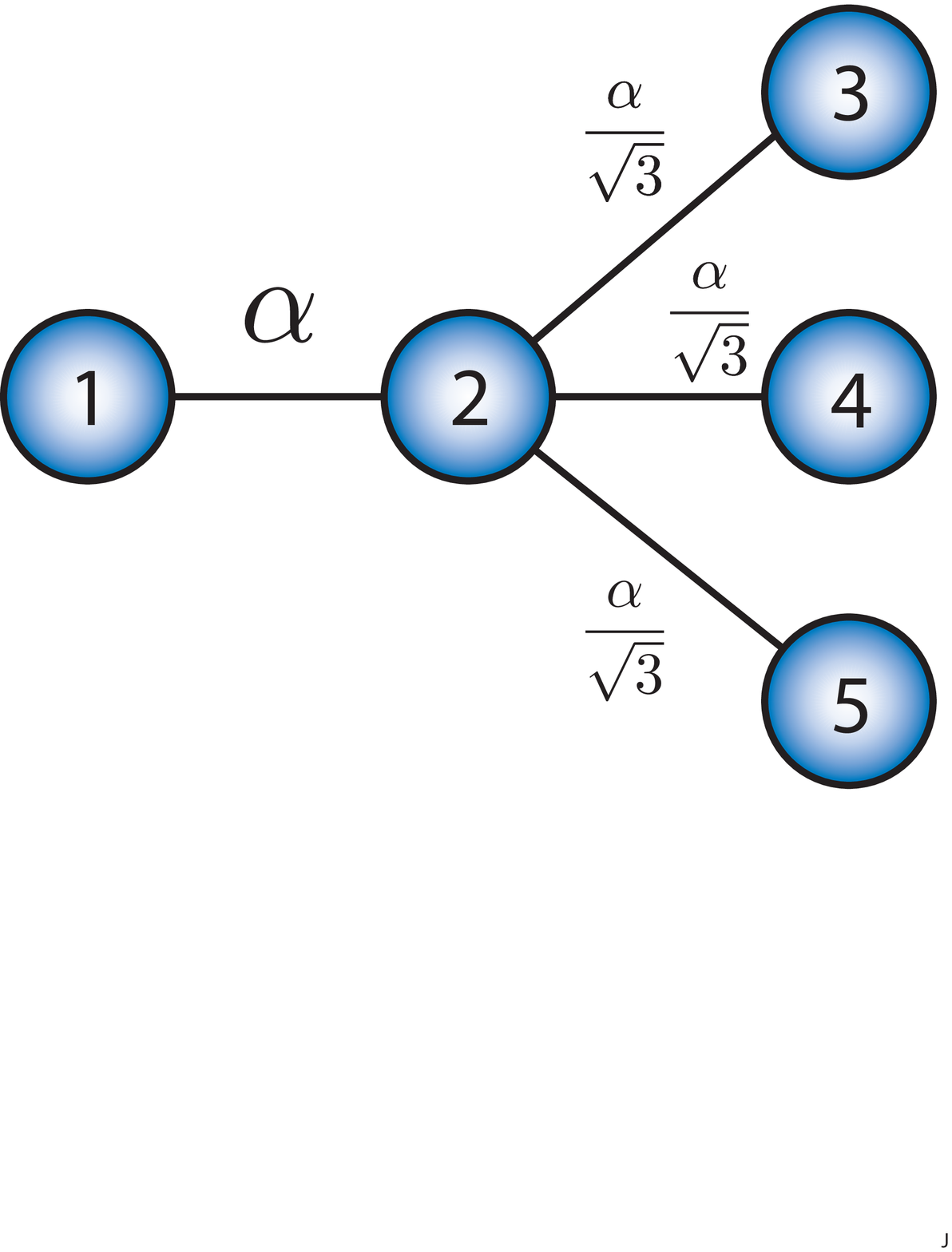}
%\vspace{-3cm}
\vspace{-3cm}
\caption{%
  A five-node, one-to-three branched spin chain. The couplings are
  chosen to effect perfect state transfer in the 1D equivalent
  three-node chain, thus creating a tripartite W-state in this system.}
  \ec\label{fig1}
\end{figure}

To give an explicit example of how these ideas extend to
multi-partite entanglement generation, we  consider the spin
chain system of figure~\ref{fig1}.
Here, using the results of \cite{chr05}, the couplings are chosen so
that the 1D equivalent spin chain follows eq.~(\ref{coupling}) and
thus perfect state transfer occurs after a time $\pi/(2 \alpha)$.
This corresponds to the creation of a tripartite W-state~\cite{dur}
\begin{equation}
\ket{W_0} =
\frac{1}{\sqrt{3}}\left(\ket{001}+\ket{010}+\ket{100}\right)
\label{Wstate}
\end{equation}
for the three output spins of the actual structure of
figure~\ref{fig1}.

Clearly the creation of the W-state is transient, as the dynamics
continues. The entanglement reappears at integer multiples of
$\pi/\alpha$ after its initial creation. However, the entanglement
can be preserved if fast (compared to $\pi/\alpha$) phase rotations
are applied to two of the output qubits at a time when the W-state
exists. From the symmetry of the Hamiltonian for the branched
system, it is clear that the two states orthogonal to
eq.~(\ref{Wstate}),
\begin{equation}
\ket{W_{\pm}} = \frac{1}{\sqrt{3}}\left(\ket{001}+e^{\pm 2i
\pi/3}\ket{010}+e^{\mp 2i \pi/3}\ket{100}\right)\;, \label{pmWstate}
\end{equation}
are eigenstates and thus play no role in the state transfer 
evolution. Therefore,
if a fast relative phase rotation of $2 \pi/3$ is applied to one
output spin, and the reverse rotation applied to another of the
state $\ket{W_0}$, the W-entanglement will be frozen. No further
non-trivial dynamics will occur. This demonstrates how multi-partite
entanglement, once created, could be preserved for future use.

A further example of this is the use of the W-state (\ref{Wstate})
to prepare a singlet. Suppose that the third output spin is 
measured in the computational basis. 
With probability 1/3 the first two output spins are projected to
$\ket{00}$ and with probability 2/3 they are projected to 
$\frac{1}{\sqrt{2}}(\ket{01}+\ket{10})$, the state being 
heralded by the measurement result. If a fast phase flip is now applied to
one of these qubits, the singlet state results. As this is a linear
superposition of $\ket{W_{+}}$ and $\ket{W_{-}}$, with no $\ket{W_0}$
component, it is frozen.

\begin{figure}[t]\bc
\includegraphics[scale=0.27]{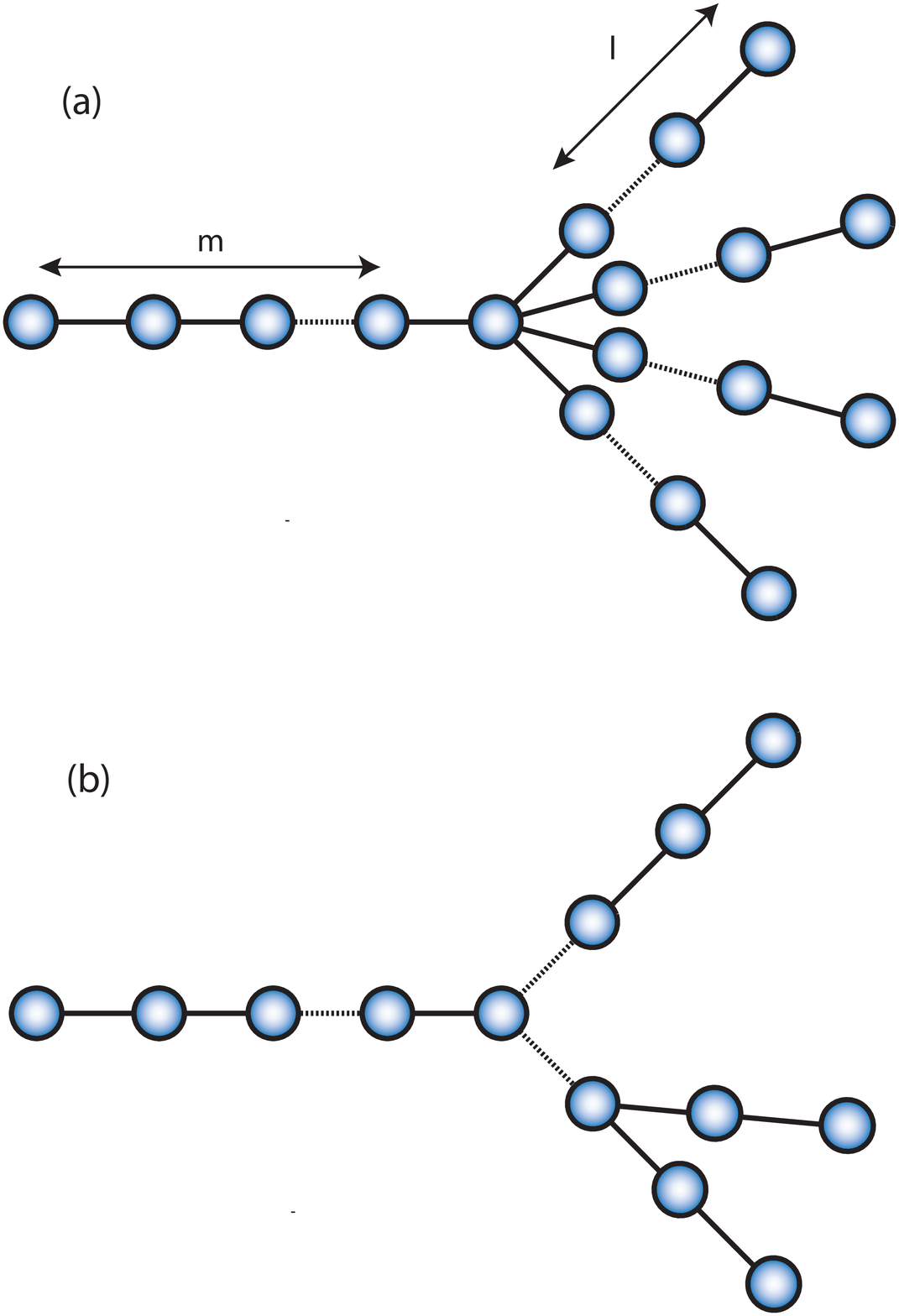}
\caption{(a) Example of a star-like family member $(m,l,l,l,l)$. (b)
Example of a bifurcation-like family member. } \label{n_out}
\ec\end{figure}

%\section{Generalization to $p$ outputs}
The branching rules we have introduced and illustrated for W-state
production can be extended to the case of the different families of
tree-like spin chain structures with a single input and $p$ output
branches. Examples of their members are represented in
figure~\ref{n_out}.

Panel (a) corresponds to a ``star-like'' family~\cite{starfam}, in which there is
just one hub and all output branches have the same length. We denote 
members of this family by $(m,l,l,...,l)$, with an
input branch of $m$ spins and $p$ output branches of $l$ spins.
Extending the discussion for one-to-two and one-to-three systems,
for a one-to-$p$ system the coupling on the outputs from the hub
spin are all pick up a factor of $1/\sqrt{p}$ compared to the
effective one-dimensional spin chain value, that would be chosen
according to eq.~(\ref{coupling}) to achieve perfect state transfer.
(Again, an analogous factor also exists for the condition of no hub
reflection for a wave packet excitation on the bosonic or
propagating fermion systems discussed
previously~\cite{perales05,yang}.) The entangled state
which can be created by using this family is a $p$-way W-state,
consisting of an excitation shared between $p$ separated sites, with
the
 {\it symmetric} form $(\ket{1_{n_2},0_{n_3},...,0_{n_p}}+\ket{0_{n_2},1_{n_3},...,0_{n_p}}+...+\ket{0_{n_2},0_{n_3},...,1_{n_p}})/\sqrt{p}$.
In real systems it is likely to be impractical to prepare $p$-way
entangled states using this approach for large values of $p$.
However, if three-dimensional physical structures can be built---
and for example there is the potential for this with quantum dot
systems---modest values of $p>2$, such as 3, 4 and 5 could be
possible. For the case $l=1$, the entanglement could again be frozen
by the application of fast relative phase factors. The phases $\theta_i$ for each output $i$ are chosen such that the resultant states are orthogonal to the dynamically produced W-type state; i.e.$ \sum_i \exp(i\theta_i) = 0$. There are always $(p-1)$ possible methods of achieving this, one of which is
to apply phases equal to the
$p$-th roots of unity to the output spins. Alternatively, for an even $p$ one can apply a $\pi$ phase to $p/2$ outputs. 

For the case $l>1$ the
freezing is not complete, although the wait time for the
entanglement to reappear is shortened due to destructive
interference preventing the spin flip from propagating back beyond
the hub, as discussed in \cite{ida07} for the bipartite
case.

Panel (b) gives an example of what we term a ``multiple
bifurcation'' structure. These structures contain more than one
hub---and for perfect transfer the branching rule must be
implemented at each. Note that for timing of the dynamical evolution
to produce complete and simultaneous excitation transfer to the
outlying sites, the number of sites along all output paths from the
initial hub must be the same. The entangled states which can be
created by using this family share the excitation between the
different separated outlying sites with different unequal weights.
For example, the structure represented in panel (b) gives the form
$\ket{1_{n_2},0_{n_3},0_{n_4}}/\sqrt{2}+\ket{0_{n_2},1_{n_3},0_{n_4}}/2+\ket{0_{n_2},0_{n_3},1_{n_4}}/2$.
With only modest branching at each hub, a bifurcating structure
potentially could be fabricated in a planar arrangement. This could
in principle give rise to a significant number of output branches,
whose outlying spins can be entangled, sharing an excitation in an
{\it asymmetric} manner, where the type of asymmetry depends on the
number of hubs and on their position.

%%% 4 %%%

\section{Conclusion}

We have illustrated how branching spin chain systems can be used to
generate and distribute multi-partite entanglement from their
natural dynamics. Clearly once it is distributed, this entanglement
could be isolated through swapping the states of the spins at the
ends of the output branches into other adjacent qubits. An
alternative approach considered in this work is the application of
 simple
single-qubit operations to the output branch end spins. Given its
simplicity, this latter approach may be particularly useful in
initial experimental investigations of such branched systems. We
illustrated this by showing how a tripartite W-state could be
created and completely frozen in a very simple branched system.
W-states are robust and immune against global dephasing~\cite{haffneret-rooset};
research is on-going into specific applications of W-states, including in teleportation~\cite{joo03} secure communication~\cite{wang07} and leadership election protocols~\cite{dhondt06}.
Generally, distributed  entanglement provides a useful resource, for example
for teleportation~\cite{tele93} or distributed quantum processing.
In contrast to the use of spin chains to propagate quantum states
from one place to another with as high a fidelity as possible, there
could be some advantage in building up a high fidelity entangled
resource ``off line''. Real systems, with their inevitable
imperfections, will almost certainly degrade transmission
fidelities, even if in principle these approach unity. Certainly,
with the ``off line'' resource approach, purification~\cite{pur96}
could be applied to build up a higher fidelity resource that can be
achieved by direct transmission, provided that the entanglement can
be preserved once it is created. High fidelity purified entanglement
could be used to transfer quantum states, or perform some other form
of quantum communication. In effect, the concept of a quantum
repeater \cite{repeater98} could be employed in a solid state, spin
chain scenario.

We comment finally that all these related effects result from the
basic dynamics of the branched spin chain systems and simply
prepared initial states. Some control is needed over the couplings to
achieve entanglement creation, distribution and preservation, but
there is certainly significant potential for branched spin chain
systems to find application in solid state quantum processing and
communication. This potential will continue to grow, as fabrication
or creation of solid state systems that can operate as spin chains
continues to progress.

\section{acknowledgement}
  BWL acknowledges support from the Royal Society through a University Research
Fellowship, the QIP IRC (www.qipirc.org, GR/S82176/01) and St Anne's College, Oxford. IDA acknowledges support from the
Dept. of Physics, University of York,
and the kind hospitality of Hewlett-Packard Labs., Bristol,
and the Dept. of Materials, University of Oxford.


\begin{thebibliography}{[1]}

\bibitem{qkd} N. Gisin, G. Ribordy, W. Tittel and H. Zbinden,
{\it Rev. Mod. Phys.} {\bf 74}, 145 (2002).

\bibitem{bose03} S. Bose, {\it Phys. Rev. Lett.}  {\bf 91}, 207901 (2003).

\bibitem{osb04} T. J. Osborne and N. Linden, {\it Phys. Rev. A} {\bf 69},
052315 (2004).

\bibitem{chr04} M. Christandl, N. Datta, A. Ekert and A. J. Landahl, {\it Phys. Rev.
Lett.} {\bf 92}, 187902 (2004).

\bibitem{chr05} M. Christandl, N. Datta, A. C. Dorlas, A. Ekert, A. Kay,
A.~J.~Landahl, {\it Phys. Rev. A} {\bf 71}, 032312 (2005).

\bibitem{dechi05} G. De Chiara, D. Rossini, S. Montangero and
R. Fazio, {\it Phys. Rev. A} {\bf 72}, 012323 (2005).

\bibitem{yun05} M.-H. Yung and
S. Bose, {\it Phys. Rev. A} {\bf 71}, 032310 (2005).

\bibitem{kar05} P. Karbach and J. Stolze, {\it Phys. Rev. A} {\bf 72}, 030301 (R) (2005).

\bibitem{bur05a} D. Burgarth and S. Bose, {\it Phys. Rev. A} {\bf 71},
052315 (2005).

\bibitem{bur05b} D. Burgarth and S. Bose, {\it New J. Phys} {\bf 7}, 135 (2005).

\bibitem{woj05} A. W\'{o}jcik, T.~Luczak, P.~Kurzynski, A.~Grudka,
T.~Gdala and M.~Bednarska, {\it Phys. Rev. A} {\bf 72}, 034303 (2005).

\bibitem{damico05} I. D'Amico, ``Relatively short spin chains as building
blocks for an all-quantum dot quantum computer architecture''  in:
{\it Semiconductor Research Trends}  (edited by K. G. Sachs, Nova Science Publishers, 2007);
({\it Preprint} cond-mat/0511470).

\bibitem{ple05} M. B. Plenio and F.L. Semiao, {\it New J. Phys.} {\bf 6}, 36 (2005).

\bibitem{har05} M. J. Hartmann, M. E. Reuter and M. B. Plenio,
{\it New J. Phys.} {\bf 8}, 94 (2005).

\bibitem{avel06} M. Avellino, A. J. Fisher and S. Bose,
{\it Phys. Rev. A} {\bf 74}, 012321 (2006).

\bibitem{gre05}  A. D. Greentree, S. J. Devitt and L. C. L.
Hollenberg, {\it Phys. Rev. A} {\bf 73}, 032319 (2006).

\bibitem{ohs07} T. Ohshima, A. Ekert, D. K. L. Oi, D. Kaslizowski and L. C.
Kwek, {\it Preprint} quant-ph/0702019.

\bibitem{plenio04} M. B. Plenio, J. Hartley and J. Eisert, {\it New J. Phys.} {\bf 7}, 73 (2004).

\bibitem{perales05} A. Perales and M. B. Plenio, {\it J. Opt. B}  {\bf 7}, S601 (2005).

\bibitem{yang} S.~Yang, Z.~Song and C.~P.~Sun, Eur. Phys. J. B {\bf 52}, 377 (2006); S.~Yang, Z.~Song and C.~P.~Sun, Front. Phys. China {\bf 1}, 1 (2007).

\bibitem{dev05} S. J. Devitt, A. D. Greentree and L. C. L.
Hollenberg, {\it Preprint} quant-ph/0511084.

\bibitem{pat05} M. Paternostro, H. McAneney and M. S. Kim, {\it Phys.
Rev. Lett.} {\bf 94}, 070501 (2005).

\bibitem{spi06} T. P. Spiller, I. D'Amico and B. W. Lovett,
{\it New J. Phys.} {\bf 9}, 20 (2007).

\bibitem{tele93} C. H. Bennett, G. Brassard, C. Crepeau, R. Jozsa,
A. Peres and W. Wootters, {\it Phys. Rev. Lett.} {\bf 70},  1895 (1993).

\bibitem{pur96} C. H. Bennett, G. Brassard, S. Popescu, B.~Schumacher,
J.~A.~Smolin, and W.~K.~Wootters, {\it Phys. Rev. Lett.} {\bf 76}, 722 (1996).
\bibitem{dur}    W. D\"ur, G. Vidal, and J. I. Cirac,{\it Phys. Rev. A} {\bf 62}, 062314 (2000)

\bibitem{starfam}Entanglement creation in a different ``star-like'' family
has been analysed by A. Hutton, and S. Bose, {\it Phys. Rev. A} {\bf 69}, 042312 (2004). 
\bibitem{ida07} I. D'Amico, B. W. Lovett and T. P. Spiller,
to be published in Phys. Rev. A Rapid (2007); I. D'Amico, B. W. Lovett and T. P. Spiller,
{\it Preprint} 	quant-ph/0702269  (2007).

%\bibitem{nie00} For our calculations we adopt the fidelity definition of
%M. A. Nielsen and I. L. Chuang, {\it Quantum Computation
%and Quantum Information} (Cambridge University Press 2000) p. 409.

%\bibitem{munro01} W.~J.~Munro, D.~F.~V.~James, A.~G.~White and P.~G.~Kwiat,
%{\it Phys.~Rev.~A} {\bf 64}, 030302 (2001).
\bibitem{haffneret-rooset}H. H\"affner, W. H\"ansel, C. F. Roos, J. Benhelm, D. Chek-al-kar, M. Chwalla, T. K\"orber, U. D. Rapol, M. Riebe, P. O. Schmidt, C. Becher, O. G\"uhne, W. D\"ur and R. Blatt, {\it Nature} {\bf 438}, 643 (2005);
C. F. Roos, M. Riebe, H. H\"affner, W. H\"ansel, J. Benhelm, G. P. T. Lancaster, C. Becher, F. Schmidt-Kaler, and R. Blatt., {\it Science} {\bf 304}, 1478(2004).

\bibitem{joo03} J. Joo, Y.-J. Park, S. Oh and J. Kim, {\it New J. Phys.} {\bf 5}, 136 (2003).

\bibitem{wang07} J. Wang, Q. Zhang and C.-j. Tang,  {\it Preprint} arXiv:quant-ph/0603144.

\bibitem{dhondt06} E. D'Hondt and P. Panangaden, {\it Quant. Inf. and Comp.} {\bf 6} 173 (2006)

\bibitem{repeater98} H.-J. Briegel, W. D\"{u}r, J. I Cirac and P. Zoller,
{\it Phys. Rev. Lett.} {\bf 81}, 5932 (1998).

\end{thebibliography}
\end{document}